\documentstyle[11pt]{article}
\begin{document}
\centerline{\large\bf Single Spin State Detection}
\centerline{\large\bf for the Kane Model of Silicon-Based Quantum Computer}
\vskip 2mm
\centerline{S.N.Molotkov and S.S.Nazin}
\vskip 2mm
\centerline{\sl\small Institute of Solid State Physics 
            of Russian Academy of Sciences}
\centerline{\sl\small Chernogolovka, Moscow district, 142432,  Russia}
\vskip 2mm

\begin{abstract}
The scheme for measurement of the state of a single spin
(or a few spin system) based on the single-electron turnstile and
injection of spin polarized electrons from magnetic metal contacts
is proposed. Applications to the recent proposal
concerning the spin gates based on a silicon matrix 
(B.Kane, Nature, {\bf 393}, 133 (1998)) are discussed.
\end{abstract}

After the discovery of efficient quantum algorithms [1] 
and a rigorous proof of the possibility of fault-tolerant
quantum computing [2], various realizations of quantum
logical gates were proposed based on cold ions
[3], nuclear magnetic resonance [4], optical schemes 
[5], semiconductor heterostructures 
[6], and Josephson effect [7]. Recently, the possibility of
fabrication of quantum gates employing a silicon matrix with
the dopant P$^{31}$ atoms was suggested [8].
The role of quantum bits is played by the nucleus and electron
spins of the P$^{31}$ atom. One of the problems in the
scheme [8] is the measurement of a single nucleus or electron
(or both) spin. The papers [9,10] consider the indirect
spin measurement employing the single-electron transistor.
However, the proposed schemes do not allow to measure the 
state of a single spin and only make it possible to measure
the different charge states of a system of nucleus and/or electron spins.

The detection of a single spin in itself is not an exotic thing.
The observation of Larmor precession of a single spin localized
on the Si(111)7$\times 7$ surface in ultra-high vacuum
with the scanning tunnelling microscope (STM) was first
reported by the IBM group (Demuth et al [11], see also [12])
ten years ago. There were also reports on the detection of
the electron paramagnetic resonance signal in the STM
current from a single spin in an organic molecule [13].
The STM with a magnetic tip was also demonstrated to be sensitive
on the atomic level to the state of single spins on the surface of 
magnetic materials [14]. The quantity which is directly measured 
in STM is the tunnel current which depends on the lateral tip position 
relative to the sample on the atomic scale, and in the case of a
magnetic tip the tunnel current has a spin-dependent component [12]
\begin{displaymath} 
I_t({\bf x})\propto \rho_c({\bf x})\rho_t\; {\bf m}_c({\bf x})\cdot{\bf m}_t, 
\end{displaymath} 
where $\rho_{c,t}({\bf x})$ are the local densities of states
on the tip apex and the sample surface, respectively,
while ${\bf m}_c({\bf x})$ and ${\bf 
m}_t$  are the local magnetizations at the sample surface
at a point ${\bf x}$ and the tip apex, respectively.
However, the steady state current measurements 
cannot be directly applied to the detection of the state
of quantum gates.

Realization of quantum gates requires the possibility
of measuring the state of the system at an arbitrary
moment of time. According to the general theory
of quantum-mechanical measurements [15--17], the most
complete description of any particular measuring procedure
which can be applied to a quantum system is given by the so-called
instrument. The instrument $T(d\lambda)$ is actually a mapping of the
set of all quantum states of the system (density matrices)
$\rho_s$ before the measurement to the system states 
(up to the normalization) just after the measurement
$\tilde{\rho}_s=T(d\lambda)\rho_s$ which gave the result in 
the interval $d\lambda$, the probability of obtaining
a measurement result in the interval $d\lambda$ being
${\rm Tr}\tilde{\rho}_s={\rm Tr}\{T(d\lambda)\rho_s\}$. 
Any instrument can be represented in the form 
$$
T(d\lambda)\rho_s=\mbox{Tr}_A\{(I_s\otimes M_A(d\lambda))
U(\rho_s\otimes\rho_A)U^{-1}\},
$$
i.e. any measurement can be realized by allowing
the studied system to interact with a suitable auxiliary system
(ancilla) prepared in a fixed initial state $\rho_A$ for some time
(so that their joint evolution is described by certain 
unitary dynamics $U$) and a subsequent measurement 
generated by a suitable identity resolution $M_A(d\lambda)$
performed on the ancilla.

Proposed below is the method for detection of the
state of a single spin (or a few spins, e.g. nucleus spin +
electron spin) based on the ``turnstile'' concept [18,19].
Explicitly present in the proposed scheme are
the preparation of the ancilla ($\rho_A$) at an arbitrary moment of time,
turning on the interaction between the ancilla and the studied system,
their joint unitary evolution, turning off the interaction at an arbitrary 
moment of time, and detection of the state of the ancilla.
Consider the following model system. Suppose that we have a system of spins
in the subsurface region, e.g. an atom with the nucleus possessing non-zero spin
and an electron localized on that atom (Fig. 1). Suppose also that
a system of tunnel-coupled quantum dots is fabricated on the surface
in such a way that the central dot is located just above the system of 
spins acting as the quantum bits (Figs. 1,2). Each dot has a single 
size-quantized level. The leftmost and the rightmost dots 
are tunnel-coupled to the magnetic metal electrodes (Figs. 1,2).

The Hamiltonian describing the interaction among the quantum dots
and between the quantum dots and metallic electrodes can be written as
\begin{equation}
H=\sum_{k,\sigma,\alpha=L,R}\varepsilon_{k\sigma\alpha}a^{+}_{k\sigma\alpha} 
a_{k\sigma\alpha}+
\sum_{\sigma}(\varepsilon_{c}c^{+}_{c\sigma}c_{c\sigma}+
\varepsilon_{L}c^{+}_{L\sigma}c_{L\sigma}+\varepsilon_{R}c^{+}_{R\sigma}c_{R\sigma})+
\end{equation}
$$
\sum_{k\sigma}(T_{kL}c^{+}_{L\sigma}a_{k\sigma L}+T_{cL}c^{+}_{c\sigma}c_{L\sigma}+
T_{cR}c^{+}_{c\sigma}c_{R\sigma}+\mbox{h.c.})+
\sum_{\sigma}(U_L n_{L\sigma}n_{L-\sigma}+U_cn_{c\sigma}n_{c-\sigma}
U_r n_{R\sigma}n_{R-\sigma}),
$$
where the first two terms describe the electron states in isolated 
electrodes and the dots, the third one represents the tunnel coupling 
between the dots and the electrodes, and the last term accounts for
the Coulomb intradot repulsion (if it is important). We assume that
the electrons in the electrodes are spin-polarized with the magnetization
vectors (${\bf n}_L$ and ${\bf n}_R$ in the left and right electrodes, 
respectively) fixed by e.g. magnetic anisotropy. If the system is placed 
in an external magnetic field, the corresponding terms should be added to the Hamiltonian.
The spin system (quantum gate) Hamiltonian, for example, for the
case of nucleus spin + the electron spin localized on it can be written as
\begin{equation}
H_s=\sum_{\sigma} \varepsilon_s c^{+}_{s\sigma} c_{s\sigma}+
g_s \mu_B c^{+}_{s\sigma} c_{s\sigma'} \mbox{\boldmath $\sigma$}_{\sigma\sigma'}
\cdot {\bf H}+ g_I\mu_B {\bf I} \cdot {\bf H}+
g_{sI}{\bf I}\cdot \mbox{\boldmath $\sigma$}_{\sigma\sigma'} 
c^{+}_{s\sigma}c_{s\sigma'}.
\end{equation}
In the external magnetic field the contribution from
the metal electrodes should also be taken into account.
The Hamiltonian describing the interaction of the spins in the quantum
gate and the electron localized in the central dot (see below)
depends on the specific geometry of the considered structure.
For example, if the wave functions of the central dot electron
and the electron localized on the subsurface center overlap,
then the Hamiltonian can be written as
\begin{equation}
H_{int}=\sum_{\sigma}(t_{sc}c^{+}_{s\sigma} c_{c\sigma}+\mbox{h.c.})+
g_{cI}{\bf I}\cdot \mbox{\boldmath $\sigma$}_{\sigma\sigma'}
c^{+}_{c\sigma}c_{c\sigma'}+
\sum_{\sigma}U_{sc}n_{c\sigma}n_{s-\sigma}.
\end{equation}
If the overlap is negligible, then only the dipole-dipole interaction 
should be retained.

The complete solution of the problem of finding
the temporal evolution of the considered system is a difficult task.
To go one step further, we shall assume that the characteristic times 
of different processes occurring in our systems form a certain
hierarchy.
Let  $\tau_{res}$ be the typical time of electron tunneling  
from the central dot to the metal electrode when the energy
levels in the adjacent quantum dots are tuned to the resonance
(this time actually coincides with the time required for the electron 
to tunnel from the left (right) quantum dot to the metal electrode
through a single barrier),
$\tau_{non}$ be the characteristic tunneling time from the central
dot to the metal electrodes when the levels in the dots are detuned
from the resonance, and finally 
$\tau_{dyn}$ be the typical time of joint evolution caused by the interaction
between the electron in the central dot and the spins in the gate.
We shall assume that $\tau_{res}\ll 
\tau_{dyn}\ll \tau_{non}$.  Below we wish to take advantage
of the well known point that for the tunneling through the two
barriers (from the central dot to the metal electrodes), tuning of the 
levels into the resonance lifts the smallness associated with
the additional barrier. The characteristic time are inversely
proportional to the level width and depends on the level position;
an estimate is given by (e.g., see Ref.[19])
$$
\frac{1}{\tau(\omega)}=\gamma(\omega)=\frac{|T_{Lc}|^2\gamma_0^2}
{[\tilde{\varepsilon}_{c}(\omega)-\tilde{\varepsilon}_L(\omega)]^2+\gamma_0^2},
\quad
\gamma_0=\sum_k |T_{kL}|^2\delta(\omega-\varepsilon_{kL})=|T_{L}|^2 .
$$
Here $\gamma_0=|T_L|^2\approx |T_{cL}|^2=|T|^2$ is the bare 
tunnel transparency of the barrier between the dots and between the dots 
and the electrodes which can be assumed to be the same without any loss of 
generality. Under the resonance conditions
($\tilde{\varepsilon}_{c}(\omega_r)=\tilde{\varepsilon}_L(\omega_r)$)
$1/\tau_{res}\approx |T|^2=\gamma_0$. When the levels are detuned
by the energy larger
than the level width ($\Delta \gg 
\gamma_0$), the characteristic time becomes 
$1/\tau_{non}\approx \gamma_0(\gamma_0^2/\Delta^2)\ll 
1/\tau_{res}$.  Accounting for the Coulomb repulsion
does not change the situation qualitatively.

Let us now discuss the different stages of the measurement procedure (Fig. 2.)

a) The size-quantized levels in the dots are initially empty
(lie above the chemical potentials in the electrodes). The dashed line
shows the levels in the dots split off by the Coulomb repulsion.

b,c) The central and the left dots are subjected to
the voltage pulses with the duration 
$\tau$ such that $\tau_{res}\ll \tau\ll \tau_{dyn}$ 
and the central and left dots levels are
tuned into the resonance and pulled down
below the chemical potential
$\mu_L$ in the left electrode.  During the time of the order
of $\tau_{res}$ the levels in the left and the central dots
are filled by the electrons from the left electrode.

d) Then the voltage pulse with the duration 
$\tau_{res}\ll \tau\ll \tau_{dyn}$ is applied to the left dot
pushing its level above the chemical potential
$\mu_L$. During the time of the order of 
$\tau_{res}$ the level in the left dot becomes empty since
the electron escapes back into the left electrode.
AT the same time the level in the central dot remains filled.
On the time scale 
$<\tau_{non}$ one can assume that the electron does not remember
about the electrodes and is effectively isolated, its spin state
being determined by the left electrode state. The above described 
procedure results in the preparation at the initial moment of time
(on the time scale $\tau\ll\tau_{dyn}$ --- instantly) 
of the ancilla in the state
$\rho_A(t=0)$.  Since the density matrix for the spin-1/2 system
can always be written in the form 
$\rho=\frac{1}{2}(I+\mbox{\boldmath $\sigma$}\cdot{\bf u})$, 
the state of the electron in the central dot which tunnelled
from the left electrode is described by the density matrix
$\rho_A(t=0)=(1/2)(I+\mbox{\boldmath $\sigma$}\cdot{\bf u}_L)$, 
where ${\bf u}_L$ is the vector describing both the direction and
the magnitude of the electron spin polarization vector
of the electrons in the left electrode.

Then on the time scale $\tau_{res}\ll t \ll \tau_{non}$ one can assume
that the electron in the central dot and the spins in the gate
evolve according the joint unitary dynamics
$$
\tilde{\rho}(t)=U(t)(\rho_A(t=0)\otimes\rho_s(t=0))U^{-1}(t),\quad
U(t)=\exp{(i\int_0^t H_{int}(t')dt')}.
$$
Here $\rho_s(t=0)$ is the density matrix of the quantum gate
the time moment $t=0$.  The Hamiltonian $H_{int}$ can be easily
diagonalized since it describes a finite-dimensional system
(e.g., the joint dynamics of the electron localized in the dot 
together with the nucleus spin and the electron spin localized on it
is described by the $8\times8$ matrix).

The density matrix of the electron in the central dot
by the times moment $t$ after the joint evolution is
$$
\rho_A(t)=\mbox{Tr}_s\{U(t)(\rho_A(t=0)\otimes\rho_s(t=0))U(t)^{-1}\} =
      \frac{1}{2}(I+\mbox{\boldmath $\sigma$}\cdot{\bf u_A}(t)),
$$
where the vector ${\bf u}_A(t)$ specifies
the spin of electron localized in the central dot
by the time moment $t$ resulting from the interaction
with the spins in the quantum gate.

e) Detection of the electron state in the central dot
is performed by measuring the current flowing into the right electrode.
For that purpose the central and the right quantum dots are subjected to the 
voltage pulses of duration $\tau_1$ (Fig. 2e) similar to those
used to inject an electron from the left electrode to the central dot.
If the time $\tau_1$ is short compared with $\tau_{res}$, the probability 
of electron escape to the right electrode is proportional to 
$\tau_1$.  Since $\tau_{dyn}\gg \tau_{res}$, at the time moment 
$t$ the interaction between the ancilla and the quantum gate
is almost instantly (on the time scale characteristic of their joint
dynamics) turned off. The probability of electron escape to the
right electrode per unit time is (to within the numerical factors)
$$
\mbox{Pr}\propto |T|^2\mbox{Tr}_{sA}\{\rho_R\cdot\rho_a(t)\}=
|T|^2\mbox{Tr}_{sA}
\{(I_s\otimes\rho_R)(U(t)(\rho_s(t=0)\otimes\rho_A(t=0))U^{-1}(t))\},
$$
where $\rho_R$ is the electron density matrix in the right electrode,
$\rho_R=\frac{1}{2}(I+\mbox{\boldmath $\sigma$}\cdot{\bf u}_R)$.
Therefore, the probability of appearance of a current pulse in the
right electrode depends on the spin state of the electron in the
central quantum dot and is
\begin{equation} 
{\rm Pr}= C\tau_1 \cdot |T|^2\{1+{\bf u}_R\cdot{\bf u}_A(t)\},
\end{equation}
where $C$ is a constant (the scalar product actually arises
from the reduction to a single spin quantization axis of
the two spinors describing the electron states in the central 
dot and the right electrode [20]).

f) Finally, the voltage pulses are applied to the
left and the central quantum dots whose magnitude and
duration chosen in such a way that the electron
escapes from the central dot (if after the previous stage there 
is still an electron in the central quantum dot) to the left electrode
with unit probability.

Let the duration of the complete cycle consisting of the stages
a)--f) be $\tau_0$.  Then at fixed
$\tau_1$ the current flowing
through the system of quantum dots will be equal to 
${\rm Pr}\cdot \frac{\textstyle e}{\textstyle \tau_0}$.  
The constant $C$ in Eq.(4) can be found from
the current measurements for the case of
parallel magnetizations in the left and right electrodes
when the interaction with the gate is turned off
($t=0$), so that ${\bf 
u}_R\cdot{\bf u}_A(t) = |{\bf u}_R|\cdot|{\bf u}_L|$ (we assume 
that the magnetizations in the electrodes, $|{\bf u}_R|$ and 
$|{\bf u}_L|$), are known). 

Thus, the current measurements in the outlined scheme
allow one to determine the ancilla polarization vector
${\bf u}_A(t)$ which depends on the initial state of the 
gate $\rho_s$ before the measurement. Strictly speaking, 
finding the gate state from the measured current
requires the determination of all three components 
of the vector $u_s$ appearing in the density matrix of the gate. 
It is obvious that this can only be done if
the current measurements are preformed for at least three different 
combinations of the system parameters. For example,
one can vary the magnetization direction in both electrodes
and the duration of ancilla interaction with the gate.
However, the problem of whether or not the tunnel current behavior
as a function of the indicted parameters provides sufficient information
for the complete recovery of the vector $u_s$ should be solved separately
for each particular interaction between the quantum gate and the ancilla.

The characteristic time of the non-resonant tunneling
can be made arbitrarily large by increasing the width
of the double barrier so that it imposes no restrictions. 
The typical time of the joint quantum dynamics of the
gate and the electron in the central dot can be estimated 
as the typical time of the dynamics of an isolated gate
which is the inverse Larmor spin precession in the external 
field [8]. In the field $B\approx 100$ Gs (0.01 T) 
this time is $1/\tau_{dyn}\approx10^{6}$ Hz. The 
resonant tunneling time can well be increased up to 
$\tau_{res}\approx 10^{-9}$ s allowing to measure 
the current pulses on the times $\tau\le 
\tau_{res}$. To avoid the smearing of the Zeeman splitting, 
the temperature should not exceed 1 mK. Growth of the operational
temperature shortens $\tau_{dyn}$ and, consequently, 
$\tau_{res}$, resulting in the reduction of the time during 
which the current pulses are measured. Note also
that both the quantum dots and silicon matrix [8]
should not contain the isotopes with non-zero nuclear spin
which prevents employment of the advanced technology
of GaAs/GaAlAs materials and requires usage of the
Si/SiGe-based systems.
 
The authors are grateful to Prof. K.A.Valiev for discussion of 
obtained results. The work was supported by the Russian Fond for Basic 
Research (project No 98-02-16640) and by the Program ``Advanced
technologies and devices of nano- and microelectronics''
(project No 02.04.5.2.40.T.50), and by the Program
``Surface Atomic Structures'' (project N 1.1.99).



\begin{thebibliography}{99}
\bibitem{1} P.W.Shor, {\it Proceedings 35th Annual Symposium on 
       Foundations of Computer Science}, Santa Fe, NM, USA, 
       ed by S.Goldwasser, (IEEE Comput. Soc. Press, Los Alamitos) 
       124 (1994).  
\bibitem{2} A.Yu.Kitaev, Uspekhi Matemat. Nauk, {\bf 52}, No 6(318), 53 (1997).  
\bibitem{3} J.I.Cirac, P.Zoller, Phys. Rev. Lett., {\bf 74}, 4091 (1995).  
\bibitem{4} N.A.Gehenfeld, I.L.Chuang, Science, {\bf 275}, 350 (1997); D.G.Cory, 
       M.D.Price, T.F.Havel, Physica, {\bf D120}, 82 (1998).
\bibitem{5} Q.A.Turchette et al, Phys. Rev. Lett., {\bf 75}, 4710 (1995); 
       G.J.Milburn, Phys. Rev. Lett., {\bf 62}, 2124 (1989).
\bibitem{6} A.Barenco, D.Deutsch, A.Ekert, R.Jozsa, Phys. Rev. Lett., 
       {\bf 74}, 4083 (1995); S.N.Molotkov, Pis'ma ZhETF, {\bf 64}, 219 (1996).
\bibitem{7} A.Shnirman, G.Sch\"on, Z.Hermon, Phys. Rev. Lett., {\bf 79}, 2371 (1997);
       L.B.Ioffe, V.B.Geshkenbein, M.V.Feigel'man, A.L.Fauch\'ere, G.Blatter, 
       Nature, {\bf 398}, 679 (1999).
\bibitem{8} B.E.Kane, Nature, {\bf 393}, 133 (1998).
\bibitem{9} B.E.Kane, N.S.McAlpine, A.S.Dzurak, R.G.Clark, G.J.Milburn, He Bi Sun,
       H.Wiseman, in {\it xxx.lanl.gov/cond-mat/9903371}.
\bibitem{10}R.Vrijen, E.Yablonovich, Kang Wang, Hong Wen Jiang, A.Balandin, 
       D.DiVincenzo, in {\it xxx.lanl.gov/quant-ph/9905096}.
\bibitem{11}Y.Manassen, R.J.Hamers, J.E.Demuth, A.J.Castellano, Jr., Phys. Rev. Lett., 
       {\bf 62}, 2513 (1989).
\bibitem{12}S.N.Molotkov, Surface Science, {\bf 264}, 235 (1992); {\bf 261}, 7 (1992);
       {\bf 302}, 235 (1994); Pis'ma ZhETF, {\bf 55}, 180 (1992); {\bf 59}, 
       178 (1994).
\bibitem{13}A.W.McKinnon, M.E.Weland, in {\it Abstracts of STM'91}, (Int. Conf.
       Scanning Tunneling Microscopy, Interlaken, 1991, p.51.
\bibitem{14}I.V.Shvets, R.Wiesendanger, D.Br\"{u}gler, G.Tarach, 
       H.-J.G\"{u}ntherodt, J.M.D.Coey, J. Appl. Phys., {\bf 71}, 5496 (1992);
       R.Wiesendanger, H.-J.G\"{u}ntherodt, G.G\"{u}ntherodt, R.Ruf, 
       Phys. Rev. Lett., {\bf 65}, 247 (1990); M.W.Prins, R.Jansen, 
       H. van Kempen, Phys. Rev., {\bf B53}, 8105 (1996).
\bibitem{15}A.S.Holevo, {\it Probabilistic and Statistical Aspects of
       Quantum Theory}. North Holland Publishing Corporation, Amsterdam, 1982.
\bibitem{16}K.Kraus, {\it States, Effects and Operations}. Springer-Verlag, Berlin, 
       1983.
\bibitem{17}P.Busch, M.Grabowski, P.J.Lahti, {\it Operational Quantum
       Physics}. Springer Lecture Notes in Physics, v.{\bf 31}, 1995.
\bibitem{18}H.Grabert, M.H.Devoret, (Editors), {\it Proceeding of the NATO ASI on 
       Single Charge Tunneling}, March, 1991. (Plenum, New York, 1992).
\bibitem{19}S.N.Molotkov and S.S.Nazin, Pis'ma ZhETF, {\bf 58}, 272 (1993).
\bibitem{20}S.N.Molotkov and S.S.Nazin, ZhETF, {\bf 107}, 1232 (1995).
\end{thebibliography}
\end{document}